\documentstyle[prd,aps,floats]{revtex}
\begin{document}

\input epsf \renewcommand{\topfraction}{0.8}
\twocolumn[\hsize\textwidth\columnwidth\hsize\csname
@twocolumnfalse\endcsname

\flushright{BROWN-HET-1364, SU-GP-03/12-1\\
NSF-KITP-03-113,
MCGILL-03-15}

\title{Inflation and Brane Gases}

\author{Robert Brandenberger~$^{a}$, Damien A. Easson~$^{b,c}$ and Anupam Mazumdar~$^{d}$}

\address{$^{a}$~Department of Physics, Brown University, Providence, Rhode Island 
02912, USA\\
$^{b}$~Department of Physics, Syracuse University, Syracuse, NY 13244-1130, USA\\
$^{c}$~Kavli Institute for Theoretical Physics, University of California, Santa Barbara, CA 93106-4030, USA\\
$^{d}$~CHEP, McGill University, Montr\'eal, QC, H3A~2T8, Canada}

\date{\today}
\maketitle

\begin{abstract}
We investigate a new way of realizing a period of cosmological
inflation in the context of brane gas cosmology. It is argued
that a gas of co-dimension one branes, out of thermal
equilibrium with the rest of the matter, has an equation of
state which can - after stabilization of the dilaton - lead to
power-law inflation of the bulk. The most promising implementation
of this mechanism might be in Type IIB superstring theory, with
inflation of the three large spatial dimensions triggered by
``stabilized embedded 2-branes''. Possible applications and
problems with this proposal are discussed.\\
\vspace{.2cm}
PACS numbers: 98.80.Cq \hfill hep-th/0307043
\end{abstract}

\vskip2pc]



\section{Introduction}

There has been a lot of recent interest 
(see e.g. \cite{Eassonrev} for a review of various
approaches) in the possibility that
string theory may provide a new scenario for the very early
Universe which can address some of the cosmological questions
\cite{RHBrev} which the inflationary Universe scenario does not answer.
In particular, extended objects which appear in string theory
may play an important role, either
in terms of their impact on the bulk dynamics 
\cite{BV,ABE}, or as hyper-surfaces
on which the matter fields of the Standard Model of particle
physics are localized (see e.g. \cite{Carstenrev} for a recent
review). 

In particular, since inflation is the most successful paradigm for 
explaining the large size and the overall approximate homogeneity of the 
Universe~\cite{Guth:1980zm} and also yields the most successful theory
for generating primordial density fluctuations and cosmic 
microwave background anisotropies (see e.g. \cite{Mukhanov:1990me,Lyth}
for reviews), it is of interest to explore whether
string theory can provide new (and hopefully better motivated)
ways of achieving a phase of cosmological inflation in the very
early Universe (see e.g. \cite{Quevedo} for a recent review).
It is possible that a phase of inflation resulting from string
theory can already be seen at the level of the four 
space-time-dimensional supergravity description of the low energy
limit of the theory. However, it is also of interest to investigate
new ways of obtaining inflation which are tied to intrinsically
stringy effects. ``Brane inflation'' is one possibility,
in which the separation \cite{Dvali,Burgess1}
or the angle \cite{Kallosh} between two branes yields the inflaton
field (see also \cite{Juan,Brodie} for related approaches).
Another way of obtaining cosmological inflation is via the
mechanism of topological inflation \cite{Stephon2,Rozali}
resulting from the defects which form as a consequence of
brane-antibrane annihilation (see e.g. \cite{Sen} for the basic
physics). Yet another possibility is to have inflation in the
context of ``mirage cosmology'', where our space-time is a test
brane moving in a curved background space-time \cite{Kehagias,Stephon1,Parry:2001zg}.

In this Letter we suggest a new way of obtaining
bulk inflation from intrinsically stringy physics. We 
observe that (in the context of considering stringy matter on
a dynamical background field, and after fixing of the dilaton
field) the winding modes \footnote{As demonstrated below,
all we require is a gas of branes with a curvature radius larger than
the Hubble radius.} of a gas of co-dimension one branes 
(out of thermal equilibrium) has
an equation of state which yields power-law inflation, as long
as the separation of the branes is much smaller than the Hubble
radius \footnote{Note that an interesting but unrelated 
mechanism by which a thermal gas of winding modes of open
strings on D-branes causes inflation of the brane in a very high energy
regime of Hagedorn density was suggested in \cite{Abel}.}. 

Since the resulting accelerated expansion of space will
blow up the curvature radius of these branes relative to the
Hubble radius, in our proposed mechanism inflation will be of finite
duration (there is no graceful exit problem), since the accelerated
expansion will end once the separation of the branes becomes
comparable to the Hubble radius. After inflation ends, the Universe 
will take on a phase of non-accelerating
dynamics governed by a network of branes with curvature radius
comparable to the Hubble radius. This phase must end before
the time of nucleosynthesis in order to avoid an overabundance
problem analogous to the ``domain wall problem'' of particle
cosmology \cite{DWproblem}. We comment on various possibilities to
avoid this problem, the most promising of which appears to be
to use ``stabilized embedded walls'' (unstable 2-branes stabilized
by plasma effects) to drive inflation.

In the following, we briefly review some of the required 
background from string theory. Next, we discuss our inflationary
solution and study the dynamics after inflation. In the final
section, we discuss ways in which our mechanism could fit into
a string cosmology scenario of the early Universe, and address
some problems of the proposed mechanism.
 
\section{Background}

Besides the perturbative
string excitations, the non-perturbative spectrum of string theory 
includes $Dp$ branes \cite{polchi}, 
hyper-surfaces with $p$ space-like and one time-like dimension on which
open strings end. There are stable and unstable branes. Stable (BPS)
branes couple to gauge fields (coming from the Ramond-Ramond sector) 
of rank $(p+1)$, where  $p$ is odd for 
Type IIB string theory, and $p$ is even for Type IIA string theory. 
There also exist unstable (non-BPS) $Dp$-branes with the complementary
dimensions, i.e. $p$ odd (even) in Type IIA (B) string theory 
\cite{Sen1,Witten} (see e.g \cite{Sen2} for a review of stable and unstable 
branes in string theory). 
The instability of a non-BPS brane is manifested in terms of the 
presence of a tachyonic field in its world volume theory. 

In the context of cosmology, both stable and unstable branes may
play a role. The field theory analogs of stable branes are
topological defects. It is well known that defects can play an
important role in early Universe cosmology 
(see e.g. \cite{ShellVil,HK,RHBtoprev} for reviews). The field
theory analogs of unstable branes are embedded defects (see following 
paragraph), and it
has recently been realized that these can also be of
importance in the early Universe. Under certain circumstances,
unstable defects are stabilized by plasma effects
\cite{Nag1,Nag2} (see e.g. \cite{Zhang,CBD,BCD} for some recent 
suggestions of their role in cosmology). Similarly, unstable branes
could be stabilized in the early Universe and play a role in
cosmology. For some possible applications of unstable branes in
cosmology unrelated to our present work see e.g.
\cite{Stephon2,Anupam1,Mahbub1}. In the following, we will call
unstable branes which are stabilized by plasma effects ``stabilized
embedded branes''.

Certain conditions on the coupling of brane or bulk fields to the
order parameter describing the unstable brane must be satisfied in
order for our proposed mechanism to work. Recall \cite{Nag1} that
unstable topological defects can be stabilized if after symmetry
breaking there are gauge fields (photon field in the case of the
electroweak Z-string) which couple asymmetrically to the different
components of the order parameter (Higgs field), and these gauge
fields are still in thermal equilibrium. We assume that brane and bulk
gauge fields play the role of the photon field (in the case of the
electroweak Z-string), i.e. they are still in thermal equilibrium
after tachyon condensation, and they couple asymmetrically to the
different components of the tachyons. It remains to be worked out in
which brane scenarios these conditions are met.

The formation of both stable branes and unstable branes stabilized
by plasma effects is analogous to the formation of topological
defects in cosmology. One scenario is that a network of branes
forms after a phase transition in M-theory which occurs as the
Universe expands and the temperature drops. By arguments due to Kibble
\cite{Kibble} and Zurek \cite{Zurek}, such a phase transition will
leave behind a random hyper-surface network of branes with a micro-physical
correlation length (the correlation length gives the mean separation
and the curvature radius of the brane network). Note that if the phase
transition takes place at temperatures significantly lower than the
Planck temperature, then the correlation length is much smaller
than the Hubble length $H^{-1}$, where $H$ is the Hubble expansion
rate. In this case, the effect of the branes on the background geometry
can be analyzed in the {\it brane gas} approximation, where one averages
the matter distribution over a Hubble volume to obtain an average energy
density and pressure which determine the evolution of the background
cosmology. Note that in this context the branes are {\it out of
thermal equilibrium}. This is how topological defects emerge in
cosmology. Since the nature of non-perturbative string theory is
not known, it is not clear that brane gas cosmology admits a phase
transition of the above nature as the temperature increases. Another
possibility (explored in detail in the context of string gases on
a fixed background geometry \cite{MiTu,Deo,Atick}) is that as the
brane gas density increases, the branes reach a Hagedorn-like 
\cite{Hagedorn} phase,
in which the energy density of matter is dominated by infinite branes.
Such infinite branes, in the context of spatial manifolds which are
toroidal, are branes with non-vanishing winding number (see also
\cite{Kogan} for work on string thermodynamics in the presence of
branes).

To fix a simple example, we consider for the moment all spatial
dimensions to be toroidal, and the initial state of matter to be a
dense brane gas containing winding modes. Given such an initial 
dense network of branes out of thermal
equilibrium, the correlation length will increase as the scale factor
$a(t)$ as the Universe expands. Fluctuations on the branes with wavelength
smaller than the Hubble radius will undergo damped oscillatory motion,
whereas fluctuation modes with wavelength greater than the Hubble radius
will be frozen. Any net translational motion will also be damped out.
In the absence of efficient annihilation between modes
with opposite winding number, the brane gas will thus approach a gas
of static branes whose energy density is dominated by the straight brane
contribution~\footnote{Note that the energy density in the many other fields
emerging from string theory will redshift faster than the energy density 
of the winding modes considered here.}.

The equation of state for a gas of straight branes
(neglecting possible gauge charges on the branes) was worked out in
\cite{Boehm}. Here we briefly review the analysis. The 
space-time history of a $p$-dimensional object in a $D = d+1$ 
dimensional space-time with metric $g_{\mu \nu}$
is represented by the world sheet $x^\mu = x^\mu(\zeta^a)$, where 
$\mu = 0, \dots, D$ labels the space-time coordinates 
and $\zeta^a$ are the coordinates on the world sheet with 
$a = 0, \dots, p$. The induced metric on the $p$-brane is 
$\gamma_{ab}=g_{\mu\nu}x^\mu_{,a} x^\nu_{,b}$.
The action of the brane is the Nambu-Goto action
\begin{equation}\label{dbi}
S_p \, = \, - \tau_p \int{d^{p+1}\zeta \, \sqrt{-\gamma}}
\,,
\end{equation}
where we have assumed a constant dilaton $\varphi$, so that
\begin{equation} \label{tension}
\tau_p \, = \, \kappa e^{-\varphi}T_p
\, = \, \kappa T_p/g \, = \, \kappa (2\pi)^{-1}g^{-1}\alpha'^{-(p+1)/2} \, . 
\end{equation}
Here, $T_p$ is the brane tension, $g=e^{\left<\varphi\right>}$ is the string 
coupling and the string scale is given by $\alpha'=\ell^2_s$. We have
inserted a factor $\kappa$ which equals 1 for stable branes but can be
much smaller than 1 for stabilized embedded branes.

Now consider a gas of $Dp$ branes embedded in $d$ spatial
dimensions. The energy-momentum tensor is given by variation of the action
(\ref{dbi}) with respect to the metric $g_{\mu\nu}$
\begin{eqnarray}\label{emt}
T^{\mu\nu} \, &=& \, \frac{-2}{\sqrt{-g}}\frac{\delta S}{\delta g_{\mu\nu}} 
\nonumber \\
&=& \, \tau_p \int d^{p+1}\zeta \, \delta^{(D)}(x-x(\zeta))
\sqrt{-\gamma}\,\gamma^{ab}\partial_a x^\mu \partial_b x^\nu \,.
\end{eqnarray}
For a $Dp$-brane gas, Eq. (\ref{emt}) leads to the equation of 
state~\cite{Boehm}
\begin{equation}
\label{main}
{\cal P}_{p} =\left[\frac{p+1}{d}v^2-\frac{p}{d}\right]\rho_{p}\,,
\end{equation} 
where ${\cal P}_{p}$ is the pressure and $\rho_{p}$ is the energy density 
of the gas of $p$ branes. In the relativistic limit, $(v^2\rightarrow 1)$ 
the branes behave as a relativistic fluid with 
$w \equiv{\cal P}_{p}/\rho_{p} = (1/d)$, while in the
non-relativistic case $(v^2\rightarrow 0)$, we obtain 
\begin{equation} \label{eos}
w \, = \, -{p \over d} \, .
\end{equation}

\section{Inflation from Brane Gases}
 
We will work in the context of an initially hot, expanding Universe
of $d$ spatial dimensions.
We start our considerations with a configuration of matter 
taken to be a dense gas of branes of
all dimensions allowed in the underlying theory. The background
is described by the Einstein action. This implies that we are 
assuming the dilaton has been fixed by some mechanism.
Under the assumptions discussed in the previous section, the expansion of   
the early Universe will produce a gas of branes which are 
straight and static on micro-physical scales, and whose equation of
state is given by (\ref{eos}).

The Einstein equations for a $d+1$ dimensional Universe can be written as
\begin{eqnarray}
\label{accel}
\frac{\ddot a}{a} \, &=& \, 
-\frac{8\pi G_{d+1}}{d(d-1)}\left[(d-2)\rho_{p}+dp_{p}\right]\,,
\\
\label{Hubble}
\frac{\dot a^2}{a^2}\, &=& \, \frac{16\pi G_{d+1}}{d(d-1)}\rho_{p}\,,
\end{eqnarray}
where $G_{d+1}$ is the $d+1$ dimensional Newton constant. 

Since the correlation lengths of branes of different dimensions
will be comparable, the matter content of a brane gas will initially be
dominated by the branes with the largest value of $p$ (the
heaviest of the branes). Neglecting the
contribution of the other branes, making use of the equation of state
(\ref{eos}) and inserting this into (\ref{accel}), one can combine
(\ref{accel}) and (\ref{Hubble}) to obtain
\begin{equation}\label{sfi}
a(t) \, \propto \, t^{2/(d-p)} \, .
\end{equation}
Thus, provided the power in (\ref{sfi}) is greater than $1$, i.e.
provided that $d = p + 1$, we obtain
accelerated expansion (power-law inflation)
\footnote{Note that this result agrees with what follows from the study of
\cite{Easther} of the late time dynamics of brane
gases in M-theory (use $\gamma = 3$ and $m_1 = 0$ in their equation (40)).}. 
In three spatial dimensions, a period of inflation is driven by domain walls ($p = 2$)
\footnote{The idea of wall-dominated inflation was first made in 
\cite{DWproblem} (see also \cite{Seckel}).}.

Our inflationary scenario has a natural graceful exit mechanism:
during the period of accelerated expansion, the correlation length $\xi(t)$
of the defect network increases relative to the Hubble radius. Once
$\xi(t)$ is comparable to $H^{-1}(t) = {{d - p} \over 2} t$, the brane
gas approximation breaks down since the distribution of the branes
averaged over a Hubble expansion time becomes inhomogeneous, and inflation
will end (see the following section for arguments as to the further
cosmological evolution).

The above criterion for the end of inflation allows us to estimate the
duration of the inflationary phase. Given our initial conditions,
the energy density is dominated by the network of straight $p$ branes.
Their energy density at the initial time $t_i$ is given by
(up to a factor of order unity)
\begin{equation} \label{indens}
\rho(t_i) \, = \, \tau_p t_i^p {{t_i} \over {\xi(t_i)}} t_i^{-d} \, ,
\end{equation}
where $\tau_p$ is the rescaled tension of the $p$ brane, the first two factors
on the right hand side of the equation give the energy of a single brane
within the Hubble volume, the third factor gives the number of branes
per Hubble volume, and the last factor is the inverse of the Hubble volume.
The brane tension is given in terms of the string length $\ell_s$ and the
string coupling constant $g$ by (\ref{tension}). Making use of the
first Friedmann equation (\ref{Hubble}), and replacing the higher dimensional
gravitational constant $G$ by the corresponding fundamental gravitational
length scale $l_f$ (defined via $G = \ell_f^{d - 1}$), we obtain
(setting $p + 1 = d$ in order to obtain inflation)
\begin{equation}
H(t_i)^2 \, = \, {8 \over {d (d - 1) g}} \kappa
\bigl({{\ell_f} \over {\ell_s}} \bigr)^d {1 \over {\ell_f \xi(t_i)}} \, .
\end{equation}
The important ratio between Hubble radius and correlation length at
time $t_i$ is
\begin{equation} \label{inratio}
{{H^{-1}(t_i)} \over {\xi(t_i)}} \, = \, 
\bigl({{d (d - 1)} \over 8}\bigr)^{1/2} g^{1/2} \kappa^{-1/2} 
\bigl({{\ell_s} \over {\ell_f}}\bigr)^{d/2}
\bigl({{\ell_f} \over {\xi(t_i)}}\bigr)^{1/2} \, .
\end{equation} 

During the period of inflation, the correlation length increases as $a(t)$
whereas the Hubble radius only linearly in time. Thus, the correlation
length has caught up to the Hubble radius at a time $t_f$ (the end of
inflation) given by
\begin{equation} \label{duration}
{{t_f} \over {t_i}} \, = \, {{H^{-1}(t_i)} \over {\xi(t_i)}} \, .
\end{equation}
The ratio of scale factors at $t_f$ and $t_i$ is the square of this
expression. The number ${\cal{N}}$ of e-foldings of inflation is given by
\begin{equation} \label{efolds}
{\cal{N}} \, = \, {\rm{ln}} \bigl( \frac{a(t_{f})}{a(t_{i})} \bigr) \, .
\end{equation}
If, as an example, we assume $\xi(t_i) = \ell_s$ (motivated by the
fact that the initial separation of topological defects after a
phase transition is of the order of the inverse of the symmetry
breaking scale \cite{Kibble,RHBtoprev}), then (\ref{efolds}) gives
\begin{equation}
{\cal{N}} \, \simeq \, 
{{\rm{ln}} (g \kappa^{-1}) + (d - 1) {\rm{ln}}{{\ell_s} \over {\ell_f}}} \, .
\end{equation}
Thus, we conclude that it is possible to obtain a sufficient number
of e-foldings of inflation to solve the problems of standard big
bang cosmology \cite{Guth:1980zm} provided that 
the ratio $\ell_s / \ell_f$ is large. Working, for example, with
$g = \kappa = 1$, $d = 3$ and ${\ell_f}^{-1} = 10^{19}{\rm GeV}$, we require
${\ell_s}^{-1} \leq 10^{7}{\rm GeV}$ to obtain more than the 55
e-foldings of inflation (required for the scenario to solve the
problems of standard big bang cosmology). For stabilized embedded
defects with $\kappa^{-1} \gg 1$, the upper bound on the string scale
is less severe (i.e. higher).

Note that, from a naive point of view (treating the fluctuations
generated in our model like the fluctuations generated in a model of
power-law inflation driven by a scalar field with exponential potential),
the inflationary period discussed above is not capable
of explaining the observed density fluctuations and microwave
anisotropies, since the predicted spectral slope $n_s$ deviates too much
from a scale-invariant $n_s = 1$ one. To first order in the power
$\alpha = 2 / (d - p)$ which appears in the scale factor $a(t)$
(see (\ref{sfi})) one obtains (see e.g. \cite{plisi})
\begin{equation}
n_s - 1 \, = \, - {2 \over {\alpha}} \, ,
\end{equation}
whereas the observed spectral index is now known to be in the
range $n_s - 1 = -0.07 \pm 0.03$ \cite{Spergel}. Thus, in order
for our scenario to make contact with observed inhomogeneities,
we require a different mechanism to produce the primordial
fluctuations. We will return to this issue later.

\section{Evolution after Inflation}

Once the correlation length becomes comparable to the Hubble radius,
inflation will end. In particular, at that time the approximation
of treating matter as a homogeneous brane gas breaks down, since
the typical density fluctuations per Hubble volume, which can
be estimated via $\delta E / E$ (where $E$ is the total energy
within the Hubble volume, and $\delta E$ is the energy which a
single brane contributes), become of order unity.

Once a Hubble volume no longer contains one of the branes
responsible for the period of inflation, the dynamics of
that volume are determined by whatever matter is left
(e.g. the lower-dimensional branes). The volume will undergo
non-accelerating power-law expansion, and now the Hubble radius
will increase relative to the correlation length $\xi(t)$ of the $p = d - 1$
branes. Thus, the dynamical evolution after the end of inflation
is governed by
\begin{equation}
\xi(t) \, \sim \, t \, .
\end{equation} 
In particular, the energy of the Universe continues to be
dominated by these $p = d - 1$ branes, leading to a severe
domain wall overabundance problem. 

Since the energy density of matter other than the $p = d - 1$
branes is suppressed by a power of $e^{\cal N}$ relative to 
the energy density of the $p = d - 1$ branes, and since the
energy density of the $p = d - 1$ branes is already quite low, namely
given by a typical mass scale 
\begin{equation}
m(t_f) \, = \, \rho(t_f)^{1/(d+1)} 
\end{equation}
which (making use of (\ref{indens}) and (\ref{duration})) is of the order
\begin{eqnarray} \label{reheat}
m(t_f) \, &=& \, \bigl( {4 \over {\pi d (d-1)}} \bigr)^{1/(d+1)} g^{-2/(d+1)}
\kappa^{2/(d+1)} \\
&&{{\ell_f} \over {\ell_s}} \bigl( {{\ell_s} \over {\ell_f}} \bigr)^{2/(d+1)}
{\ell_s}^{-1} \, , \nonumber
\end{eqnarray}
there is also a potential problem of how to reheat the Universe.

Thus, it appears that in order to obtain an
acceptable late time cosmology, the $p = d - 1$ branes must decay,
both to solve the domain wall problem and to achieve reheating.

As mentioned at the end of the previous section, the adiabatic
fluctuations generated by our model of inflation have the wrong
spectral index, namely $n_s = 0$, and thus cannot explain the
observed large-scale structure and microwave anisotropies. Thus,
the model needs additional ingredients in order to
successfully connect with present observations. Having said this,
it is important to realize that the fluctuations generated in our
model will {\it not} be equivalent to the fluctuations generated in
a scalar field toy model which yields the same expansion rate,
the reason being that branes (like defects) generated iso-curvature
fluctuations. As long as the branes have not decayed, a spectrum
of curvature fluctuations which is scale-invariant when measured
at Hubble radius crossing will result \cite{TB86}. At the time
of the brane decay, these primordial iso-curvature fluctuations
will contribute to an adiabatic mode. The details of this process
remain to be studied. In analogy to what can be achieved \cite{Bozza} in
the Pre-Big-Bang \cite{PBB} scenario, where the primordial adiabatic
spectrum has the wrong index (namely $n_s = 4$ \cite{BGGMV}) 
it is possible that other matter fields obtain a scale-invariant
spectrum of iso-curvature modes which upon the decay of these fields
yields an adiabatic mode via the curvaton mechanism 
\cite{Mollerach,Wands,Sloth,JP}. Since the spectrum of our primordial
adiabatic mode is red, it is likely that any additional mechanism like
the two possibilities we just mentioned will dominate the spectrum on
scales of present cosmological interest.

\section{Discussion}

We have proposed a new mechanism for obtaining bulk inflation in
the context of string cosmology. The mechanism is based on a gas
of $p = d - 1$ branes causing power-law inflation of the 
$d$ space-dimensional bulk in
which the value of the dilaton is fixed. The duration
of inflation is automatically finite in our scenario. However,
the final state is faced with both a domain wall problem and a
reheating problem.

The first issue we wish to focus on in this section is how the 
domain wall and reheating problems can be resolved. Our idea is
that this will occur via the decay of the branes which are
responsible for inflation. The decay will release the brane
energy into ordinary matter and cause the ``temperature'' (defined
here as the fourth root of the energy density) of the Universe
to take on the value given in (\ref{reheat}). 

How can brane decay occur? If the branes which are driving inflation
are stabilized embedded branes, then the branes will automatically
decay once the temperature of the plasma has dropped sufficiently
low (see \cite{Nag1,Nag2}). If the branes are stable, the problem
is more severe. One possibility is that eventually the
branes and anti branes will annihilate, leaving a state with no long
branes. This possibility, although it works for topologically
stable branes as well as for branes which are dynamically
stable but not topologically stable (no conserved quantum number), 
seems hard to reconcile with the topological
arguments for the existence of long branes (arguments derived in
the context of topological defects in field theory - see the
reviews \cite{ShellVil,HK,RHBtoprev}). Another possibility (suggested
first by Seckel \cite{Seckel}) is that holes in the branes nucleate
and ``eat up'' the branes at the speed of light. Since this occurs
in the post-inflationary phase, there is no causality obstruction
to the local decay of the network of branes within one Hubble volume.
However, the local decay is only possible if the walls are not
protected by a topological quantum number.

It is important that brane decay occurs sufficiently early such that
the Universe has time to homogenize and thermalize sufficiently before the time
of recombination (see e.g. \cite{BCD} for a study of the corresponding
constraint on stabilized embedded topological defects). 
There are also possible constraints from inhomogeneous
nucleosynthesis.

The second question addressed in this section is how the mechanism
suggested in this Letter might fit into a coherent picture of early
Universe string cosmology, in particular whether there is any
way to connect the proposal made here with the ideas of \cite{BV,ABE}
(see also the more recent work of \cite{EJG,Easson:2001fy})
on how the dynamics of brane winding modes can create a hierarchy
of scales between the three spatial dimensions we see and the other
spatial dimensions. 

In the following, we will take three different starting points 
corresponding to different corners of the M-theory moduli space
and discuss how our proposal might fit in. First, consider {\it 11-d
supergravity}. This theory contains stable M2 and M5 branes. In
this context, the dilaton is fixed and the background equations
are the Einstein equations. Easther et al. \cite{Easther} have
recently studied the effects of winding modes of M2 branes (on 
toroidal Universe where space is $T^{10}$) on the dynamics of a
homogeneous but anisotropic cosmology and have found that, given
initial conditions for which the radii of all the tori are
comparable, the
dynamics favors three spatial dimensions becoming larger than the
other ones (which we call ``internal''). Implicitly, the
assumption was made that the M5 branes have all annihilated. 
We will make a stronger assumption, fixing the radion 
fields corresponding to the internal dimensions, and start with
a dense gas of M2 branes. The M2 branes which wrap two of the large 
spatial dimensions can give rise to inflation by the mechanism
we are proposing (M5 branes wrapping three internal dimensions
would have the same effect). Since the M2 branes are stable,
this scenario has a serious domain wall problem (which, as
mentioned at the beginning of this section, can be addressed
successfully).

As suggested by \cite{Stephon3}, it is possible that brane 
intersections play an important role in M-theory cosmology.
This might also lead to new possibilities for brane gas inflation.
We leave an investigation of this possibility to future work.

As an orthogonal starting point we consider {\it Type IIA} string
theory. Again, we fix the dilaton and the radion fields by hand.
This theory contains stable D2 branes which will generated inflation
according to our proposed mechanism. Higher dimensional stable branes
for which all but two cycles wrap internal dimensions will contribute
to inflation in an analogous way. As in the previous example, we are
faced with a serious domain wall problem.

Turning now to {\it Type IIB} string theory, we first again assume
that the dilaton and radion fields have been fixed by some mechanism
unrelated to our scenario. In this case, there are stable D1 branes
and unstable 2 branes. The unstable 2 branes, provided that they
are stabilized by plasma effects early on, can drive a period of
inflation by our mechanism. They will decay at late times, and will
thus not lead to a domain wall problem. The stable D1 branes will
take on a scaling solution like cosmic strings (without dominating
the energy density of the Universe) \cite{ShellVil,HK,RHBtoprev},
and will lead to predictions for fluctuations which are in principle
testable. Specifically, these strings yield line discontinuities in
the cosmic microwave temperature maps \cite{Kaiser}. In order not
to conflict with present observational bounds (the most stringent
bounds come from the location and narrowness of the first acoustic
peak in the spectrum of microwave anisotropies (see e.g. \cite{CSbounds}
for a recent review on the confrontation between topological defect
models and observational constraints),
the string scale must be sufficiently low such that the string tension
is less than about $10^{28} {\rm Gev}^2$.

In the context of Type IIB string theory we can try to merge the
ideas of creating a hierarchy in the scales of spatial dimensions
\cite{BV,ABE} with the mechanism for inflation proposed here. We
will make use of the fact that odd-dimensional branes are stable
and heavy, but even-dimensional branes are unstable. If they are
stabilized by plasma effects, it is likely that the core will undergo
a ``core phase transition'' \cite{CPT} analogous to what occurs for 
stabilized embedded defects in field theory \cite{Nag1,CBD,Nag2}. In
this case, the effective tension is much smaller than it would be for
a stable brane of the same dimension (this was our motivation for
introducing the factor $\kappa$ in (\ref{tension})). Let us now
begin the time evolution with all spatial dimensions of the same
size, and the dilaton free. As discussed in \cite{ABE}, all
stable higher dimensional branes annihilate, leaving behind the
stable D1 branes. These will annihilate only in three spatial
dimensions, allowing these to grow, and keeping the others small. Note
that if the stabilized embedded branes are much lighter than the
stable branes, there still will be stabilized embedded 2 branes
after the hierarchy of spatial dimensions develops. Since in the
three large dimensions there are no more stable branes present,
the stabilized embedded 2 branes will now dominate the dynamics,
leading to inflation of these dimensions, and without creating
a domain wall problem. 

To summarize the discussion of the previous paragraph, it appears
that in the context of Type IIB it is possible to connect the
speculations of \cite{BV,ABE} on the origin of the hierarchy
between the sizes of spatial dimensions with late time cosmology.
Key is the mechanism proposed in this Letter, namely inflation
triggered by a gas of stabilized embedded membranes.

\section*{Acknowledgments}

We wish to thank Cliff Burgess and Rob Myers for useful discussions.
At Brown, this work has been supported in part by the U.S. Department
of Energy under Contract DE-FG02-91ER40688, TASK A. D.E. is supported in part by NSF-PHY-0094122 and
other funds from Syracuse University and NSF-PHY-99-07949 (KITP, UC Santa Barbara). 
A.M. acknowledges support from NSERC. A. M. is a CITA National fellow.

\vskip10pt




\begin{references}


\bibitem{Eassonrev}
D.~A.~Easson,
Int.\ J.\ Mod.\ Phys.\ A {\bf 16}, 4803 (2001)
[arXiv:hep-th/0003086].

\bibitem{RHBrev}
R.~H.~Brandenberger,
arXiv:hep-ph/9910410.

\bibitem{BV}
R.~H.~Brandenberger and C.~Vafa,
Nucl.\ Phys.\ B {\bf 316}, 391 (1989).

\bibitem{ABE}
S.~Alexander, R.~H.~Brandenberger and D.~Easson,
Phys.\ Rev.\ D {\bf 62}, 103509 (2000)
[arXiv:hep-th/0005212];\\
R.~Brandenberger, D.~A.~Easson and D.~Kimberly,
Nucl.\ Phys.\ B {\bf 623}, 421 (2002)
[arXiv:hep-th/0109165].

\bibitem{Carstenrev}
P.~Brax and C.~van de Bruck,
Class.\ Quant.\ Grav.\  {\bf 20}, R201 (2003)
[arXiv:hep-th/0303095].

\bibitem{Guth:1980zm}
A.~H.~Guth,
Phys.\ Rev.\ D {\bf 23}, 347 (1981).

\bibitem{Mukhanov:1990me}
V.~F.~Mukhanov, H.~A.~Feldman and R.~H.~Brandenberger,
Phys.\ Rept.\  {\bf 215}, 203 (1992).

\bibitem{Lyth}
D.~H.~Lyth and A.~Riotto,
Phys.\ Rept.\  {\bf 314}, 1 (1999)
[arXiv:hep-ph/9807278].

\bibitem{Quevedo}
F.~Quevedo,
Class.\ Quant.\ Grav.\  {\bf 19}, 5721 (2002)
[arXiv:hep-th/0210292].

\bibitem{Dvali}
G.~R.~Dvali and S.~H.~Tye,
Phys.\ Lett.\ B {\bf 450}, 72 (1999)
[arXiv:hep-ph/9812483].

\bibitem{Burgess1}
C.~P.~Burgess, M.~Majumdar, D.~Nolte, F.~Quevedo, G.~Rajesh and R.~J.~Zhang,
JHEP {\bf 0107}, 047 (2001)
[arXiv:hep-th/0105204]; \\
D.~Choudhury, D.~Ghoshal, D.~P.~Jatkar and S.~Panda,
arXiv:hep-th/0305104.

\bibitem{Kallosh}
C.~Herdeiro, S.~Hirano and R.~Kallosh,
JHEP {\bf 0112}, 027 (2001)
[arXiv:hep-th/0110271]; \\
J.~Garcia-Bellido, R.~Rabadan and F.~Zamora,
JHEP {\bf 0201}, 036 (2002)
[arXiv:hep-th/0112147]; \\
M.~Gomez-Reino and I.~Zavala,
JHEP {\bf 0209}, 020 (2002)
[arXiv:hep-th/0207278].

\bibitem{Juan}
K.~Dasgupta, C.~Herdeiro, S.~Hirano and R.~Kallosh,
Phys.\ Rev.\ D {\bf 65}, 126002 (2002)
[arXiv:hep-th/0203019].

\bibitem{Brodie}
J.~H.~Brodie and D.~A.~Easson,
arXiv:hep-th/0301138.

\bibitem{Stephon2}
S.~H.~Alexander,
Phys.\ Rev.\ D {\bf 65}, 023507 (2002)
[arXiv:hep-th/0105032].

\bibitem{Rozali}
S.~Alexander, R.~Brandenberger and M.~Rozali,
arXiv:hep-th/0302160.

\bibitem{Sen}
A.~Sen,
JHEP {\bf 9808}, 012 (1998)
[arXiv:hep-th/9805170].

\bibitem{Kehagias}
A.~Kehagias and E.~Kiritsis,
JHEP {\bf 9911}, 022 (1999)
[arXiv:hep-th/9910174].

\bibitem{Stephon1}
S.~H.~Alexander,
JHEP {\bf 0011}, 017 (2000)
[arXiv:hep-th/9912037].

\bibitem{Parry:2001zg}
M.~F.~Parry and D.~A.~Steer,
JHEP {\bf 0202}, 032 (2002)
[arXiv:hep-ph/0109207]; \\
E.~Kiritsis, G.~Kofinas, N.~Tetradis, T.~N.~Tomaras and V.~Zarikas,
JHEP {\bf 0302}, 035 (2003)
[arXiv:hep-th/0207060].

\bibitem{Abel}
S.~A.~Abel, K.~Freese and I.~I.~Kogan,
JHEP {\bf 0101}, 039 (2001)
[arXiv:hep-th/0005028].

\bibitem{DWproblem}
Y.~B.~Zeldovich, I.~Y.~Kobzarev and L.~B.~Okun,
Zh.\ Eksp.\ Teor.\ Fiz.\  {\bf 67}, 3 (1974)
[Sov.\ Phys.\ JETP {\bf 40}, 1 (1974)].

\bibitem{polchi}
J. Polchinski, {\em String Theory, Vols. I and II}, 
(Cambridge University Press, Cambridge, 1998).

\bibitem{Sen1}
A.~Sen,
JHEP {\bf 9812}, 021 (1998)
[arXiv:hep-th/9812031].

\bibitem{Witten}
E.~Witten,
JHEP {\bf 9812}, 019 (1998)
[arXiv:hep-th/9810188].

\bibitem{Sen2}
A.~Sen,
arXiv:hep-th/9904207.

\bibitem{ShellVil} A. Vilenkin and E.P.S. Shellard;
\textit{Cosmic Strings and Other Topological Defects},
(Cambridge Univ. Press, Cambridge, 1994).

\bibitem{HK}
M.~B.~Hindmarsh and T.~W.~Kibble,
Rept.\ Prog.\ Phys.\  {\bf 58}, 477 (1995)
[arXiv:hep-ph/9411342].

\bibitem{RHBtoprev}
R.~H.~Brandenberger,
Int.\ J.\ Mod.\ Phys.\ A {\bf 9}, 2117 (1994)
[arXiv:astro-ph/9310041].

\bibitem{Nag1}
M.~Nagasawa and R.~H.~Brandenberger,
Phys.\ Lett.\ B {\bf 467}, 205 (1999)
[arXiv:hep-ph/9904261].

\bibitem{Nag2}
M.~Nagasawa and R.~Brandenberger,
Phys.\ Rev.\ D {\bf 67}, 043504 (2003)
[arXiv:hep-ph/0207246].

\bibitem{Zhang}
R.~H.~Brandenberger and X.~m.~Zhang,
Phys.\ Rev.\ D {\bf 59}, 081301 (1999)
[arXiv:hep-ph/9808306].

\bibitem{CBD}
B.~Carter, R.~H.~Brandenberger and A.~C.~Davis,
Phys.\ Rev.\ D {\bf 65}, 103520 (2002)
[arXiv:hep-ph/0201155].

\bibitem{BCD}
R.~H.~Brandenberger, B.~Carter and A.~C.~Davis,
Phys.\ Lett.\ B {\bf 534}, 1 (2002)
[arXiv:hep-ph/0202168].

\bibitem{Anupam1}
A.~Mazumdar, S.~Panda and A.~Perez-Lorenzana,
Nucl.\ Phys.\ B {\bf 614}, 101 (2001)
[arXiv:hep-ph/0107058].

\bibitem{Mahbub1}
M.~Majumdar and A.~C.~Davis,
arXiv:hep-th/0304226.

\bibitem{Kibble}
T.~W.~Kibble,
Acta Phys.\ Polon.\ B {\bf 13}, 723 (1982).

\bibitem{Zurek}
W.~H.~Zurek,
Acta Phys.\ Polon.\ B {\bf 24}, 1301 (1993).

\bibitem{MiTu}
D.~Mitchell and N.~Turok,
Phys.\ Rev.\ Lett.\  {\bf 58}, 1577 (1987); \\
D.~Mitchell and N.~Turok,
Nucl.\ Phys.\ B {\bf 294}, 1138 (1987).

\bibitem{Deo}
N.~Deo, S.~Jain and C.~I.~Tan,
Phys.\ Rev.\ D {\bf 40}, 2626 (1989).


\bibitem{Atick}
J.~J.~Atick and E.~Witten,
Nucl.\ Phys.\ B {\bf 310}, 291 (1988).

\bibitem{Hagedorn}
R.~Hagedorn,
Nuovo Cim.\ Suppl.\  {\bf 3}, 147 (1965).

\bibitem{Kogan}
S.~A.~Abel, J.~L.~Barbon, I.~I.~Kogan and E.~Rabinovici,
JHEP {\bf 9904}, 015 (1999)
[arXiv:hep-th/9902058].

\bibitem{Boehm}
T.~Boehm and R.~Brandenberger,
arXiv:hep-th/0208188.

\bibitem{Easther}
R.~Easther, B.~R.~Greene, M.~G.~Jackson and D.~Kabat,
Phys.\ Rev.\ D {\bf 67}, 123501 (2003)
[arXiv:hep-th/0211124].

\bibitem{Seckel}
D.~Seckel,
{\it Prepared for Inner Space/ Outer Space: Conference on Physics at the Interface of Astrophysics / Cosmology and Particle Physics, Batavia,
Illinois, 2-5 May 1984}

\bibitem{plisi}
C.~A.~Terrero-Escalante,
arXiv:astro-ph/0204066.

\bibitem{Spergel} 
D.~N.~Spergel {\it et al.},
arXiv:astro-ph/0302209.

\bibitem{TB86}
N.~Turok and R.~H.~Brandenberger,
Phys.\ Rev.\ D {\bf 33}, 2175 (1986).

\bibitem{Bozza} 
V.~Bozza, M.~Gasperini, M.~Giovannini and G. Veneziano,
Phys.\ Lett.\ B {\bf 543}, 14 (2002)
[arXiv:hep-ph/0206131].

\bibitem{PBB} M. Gasperini and G. Veneziano,
M.~Gasperini and G.~Veneziano,
Astropart.\ Phys.\  {\bf 1}, 317 (1993)
[arXiv:hep-th/9211021].

\bibitem{BGGMV}
R.~Brustein, M.~Gasperini, M.~Giovannini, V.~F.~Mukhanov and G.~Veneziano,
Phys.\ Rev.\ D {\bf 51}, 6744 (1995)
[arXiv:hep-th/9501066].

\bibitem{Mollerach} S. Mollerach
S.~Mollerach,
Phys.\ Rev.\ D {\bf 42}, 313 (1990).

\bibitem{Wands}
D.~H.~Lyth and D.~Wands,
Phys.\ Lett.\ B {\bf 524}, 5 (2002)
[arXiv:hep-ph/0110002].

\bibitem{Sloth}
K.~Enqvist and M.~S.~Sloth,
Nucl.\ Phys.\ B {\bf 626}, 395 (2002)
[arXiv:hep-ph/0109214].

\bibitem{JP}
T.~Moroi and T.~Takahashi,
Phys.\ Lett.\ B {\bf 522}, 215 (2001)
[Erratum-ibid.\ B {\bf 539}, 303 (2002)]
[arXiv:hep-ph/0110096].

\bibitem{EJG}
R.~Easther, B.~R.~Greene and M.~G.~Jackson,
Phys.\ Rev.\ D {\bf 66}, 023502 (2002)
[arXiv:hep-th/0204099].

\bibitem{Easson:2001fy}
D.~A.~Easson,
to appear in Int. J. Mod. Phys. A, 
arXiv:hep-th/0110225.

\bibitem{Stephon3}
S.~H.~Alexander,
arXiv:hep-th/0212151.

\bibitem{Kaiser}
N.~Kaiser and A.~Stebbins,
Nature {\bf 310}, 391 (1984).

\bibitem{CSbounds}
R.~Durrer, M.~Kunz and A.~Melchiorri,
Phys.\ Rept.\  {\bf 364}, 1 (2002)
[arXiv:astro-ph/0110348].

\bibitem{CPT}
M.~Axenides and L.~Perivolaropoulos,
Phys.\ Rev.\ D {\bf 56}, 1973 (1997)
[arXiv:hep-ph/9702221]; \\
M.~Axenides, L.~Perivolaropoulos and M.~Trodden,
Phys.\ Rev.\ D {\bf 58}, 083505 (1998)
[arXiv:hep-ph/9801232].
\end{references}
\end{document}